\documentclass[conference,a4paper]{IEEEtran}
\IEEEoverridecommandlockouts
\usepackage{cite}
\usepackage{amsmath,amssymb,amsfonts}
\usepackage{algorithmic}
\usepackage[pdftex]{graphicx}
\usepackage{textcomp}
\usepackage{caption}
\usepackage{subcaption}
\usepackage{booktabs}
\usepackage[dvipsnames]{xcolor}
\usepackage{tikz}
\usetikzlibrary{positioning}
\usepackage[pscoord]{eso-pic}
\usepackage{framed}

\usepackage{etoolbox}
\usepackage{framed}
\newcommand{\tmpcomment}[3]{{\emph{\textcolor{#1}{#2: #3}}}}
\newcommand{\sebastian}[1]{\tmpcomment{orange}{Sebastian}{#1}}
\newcommand{\babak}[1]{\tmpcomment{purple}{Babak}{#1}}

\newtoggle{hidecomments}
\toggletrue{hidecomments}

\iftoggle{hidecomments}{
	\renewcommand{\tmpcomment}[3]{}
}{
}

\usepackage{url} 

\newcommand{\accessedDate}{March~2020}
\newcommand{\urlfootnote}[1]{\footnote{\url{#1} Accessed \accessedDate}}

\newcommand{\placetextbox}[3]{
  \AddToShipoutPictureFG*{
    \put(\LenToUnit{#1\paperwidth},\LenToUnit{#2\paperheight}){\vtop{{\null}
        \framebox[\textwidth]{\parbox{\dimexpr\textwidth-2\fboxsep-2\fboxrule}{\footnotesize{#3}}}  
    
    }}%
  }%
}%

\def\BibTeX{{\rm B\kern-.05em{\sc i\kern-.025em b}\kern-.08em
    T\kern-.1667em\lower.7ex\hbox{E}\kern-.125emX}}

\begin{document}


\title{Application of Just-Noticeable Difference in Quality as Environment Suitability Test for Crowdsourcing Speech Quality Assessment Task}

\author{\IEEEauthorblockN{Babak Naderi}
\IEEEauthorblockA{\textit{Quality and Usability Lab} \\
\textit{Technische Universit\"att Berlin, DFKI}\\
Berlin, Germany \\
babak.naderi@tu-berlin.de}
\and
\IEEEauthorblockN{Sebastian M\"oller}
\IEEEauthorblockA{\textit{Quality and Usability Lab} \\
\textit{Technische Universit\"att Berlin, DFKI}\\
Berlin, Germany \\
sebastian.moeller@tu-berlin.de}
}

\maketitle

\begin{abstract}
Crowdsourcing micro-task platforms facilitate subjective media quality assessment by providing access to a highly scaleable, geographically distributed and demographically diverse pool of crowd workers. Those workers participate in the experiment remotely from their own working environment, using their own hardware.  
In the case of speech quality assessment, preliminary work showed that environmental noise at the listener’s side  and the listening device (loudspeaker or headphone) significantly affect perceived quality, and consequently the reliability and validity of subjective ratings.
As a consequence, ITU-T Rec. P.808 specifies requirements for the listening environment of crowd workers when assessing speech quality. In this paper, we propose a new Just Noticeable Difference of Quality (JNDQ) test as a remote screening method for assessing the suitability of the work environment for participating in speech quality assessment tasks. In a laboratory experiment, participants performed this JNDQ test with different listening devices in different listening environments, including a silent room according to ITU-T Rec. P.800 and a simulated  background noise scenario. Results show a significant impact of the environment and the listening device on the JNDQ threshold. Thus, the combination of listening device and background noise needs to be screened in a crowdsourcing speech quality test. We propose a minimum threshold of our JNDQ test as an easily applicable screening method for this purpose.

\end{abstract}



\section{Introduction}
\placetextbox{0.07}{0.1}{©2020 IEEE. Personal use of this material is permitted. Permission from IEEE must be obtained for all other uses, in any current or future media, including reprinting/republishing this material for advertising or promotional purposes, creating new collective works, for resale or redistribution to servers or lists, or reuse of any copyrighted component of this work in other works. This paper has been accepted for publication in the 2020 Twelfth International Conference on Quality of Multimedia Experience (QoMEX).}


Speech quality assessment has been a common problem addressed since the early days of telephony, and has also been subject for research since then. Because of the subjective nature of the concept of quality\cite{le2012qualinet,Jekosch05}, speech quality assessment is commonly carried out by human test participants who are either instructed to listen to short stretches of speech and afterwards rate perceived quality on one or several rating scales (listening-only tests), or to hold a conversation over a telecommunication system and afterwards rate the quality of the conversation, again on one or several rating scales (conversation test). The devices used for listening or conversing are commonly well-defined and controlled, in order to not provide a further negative impact on the speech samples to-be-judged. In addition, such tests are commonly carried out in a quiet laboratory environment in order to not interfere with the listening and rating task, or to not incite conversation partners to rise their voice to fight against noise in their own environment. ITU-T Rec. P.800 \cite{ITU-P800} provides requirements to the design and set-up of such assessments.

Whereas laboratory tests help to keep the measurement process clean from external sources which might disturb the measurement process (thus increasing the reliability of the measurement), they lack realism, in that the devices used and the test environments do not reflect typical usage situations. To fight this problem, and to provide a complementary efficient method to speech quality assessment, crowdsourcing  experiments have been designed \cite{ITU-P808}. Such tests make use of online workers (so-called crowdworkers) who carry out assessment tasks on a dedicated crowdsourcing platform, and get paid for their service. In crowdsourcing, the physical environment where the experiment is carried out differs between participants, and there is no control over it. At most, crowdworkers can be instructed to execute the test in a specific environment (which however will not be equivalent to a laboratory environment with respect to ambient noise, reverberation, etc.). However, it is possible to monitor the environmental conditions either by requesting crowdworkers to participate in a pre-test, or by analyzing incoming sensory signals (e.g. microphone signal).
  
Mostly, crowdworkers work at home \cite{naderi2018motivation,jimenez2019background} where they might be subject to two types of noise: one is constant noise resulting from machines or environment traffic scenes, and the other is a periodic, melodic or voiced type of noise like from TV, music, radio, or people talking \cite{jimenez2019background}. Preliminary work showed that the type of listening device (loudspeaker or headphone)\cite{ribeiro2011crowdsourcing} and environmental noise at the listener’s side \cite{naderi2018speech} affects the outcome of a speech quality test. This paper focuses on aforementioned background noises at the listener-side, and aims at identifying a suitable method for assessing the environmental conditions during online speech quality assessment test. For this purpose, a modified Just-Noticeable Difference (JND) test is developed which evaluates the environment prior to the quality assessment task. To develop the test, the JND of speech quality perceivable by an average participant with normal hearing ability will be measured in a laboratory study. Then, in the crowdsourcing study to be evaluated, the test checks whether a crowdworker can reach the same or a finer JND level. If this is fulfilled, it is assumed that the participant’s environment (and test setup) is appropriate for performing a speech quality assessment test.

The paper is organized as follows: In Section II, we describe a JND test for quality assessment, and its modified version which is proposed for the crowdsourcing paradigm. We further outline a laboratory experiment and a complementing crowdsourcing test which will serve as a basis for judging the performance of the proposed monitoring method. The results of these tests are analyzed in Section III, by checking the impact of the JND test in relation to the crowdsourcing speech quality assessment outcome. A discussion and proposals for future work conclude the paper in Section IV.

\section{Method}
In this section we first describe the just-noticeable difference of quality test, its application as an environment suitability test and its modified version to be used in crowdsourcing tests. Finally, we explain the laboratory and crowdsourcing experiments we conducted to find proper settings for the JNDQ test.

\begin{figure*}[htbp]
\centerline{\includegraphics[width=0.8\textwidth]{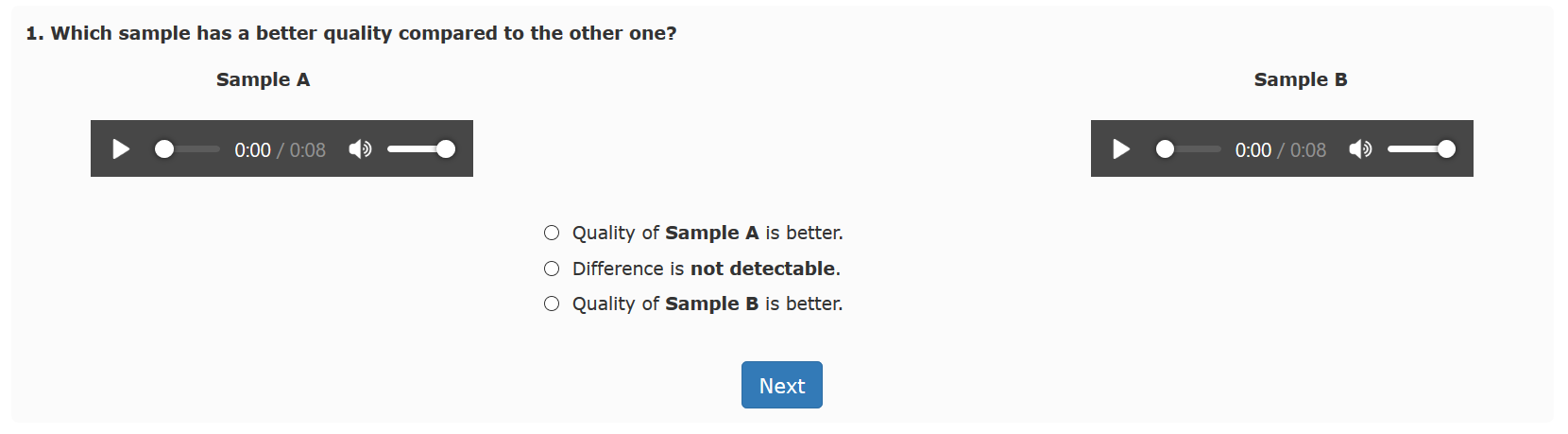}}
\caption{A sample question from the Just-Noticeable Difference of Quality Test.}
\label{fig_pair_cmp}
\end{figure*}

\subsection{Just-Noticeable Difference of Quality (JNDQ) Test}
The Just-Noticeable Difference (JND) refers to the minimum difference between two stimuli that leads to a change in experience \cite{treutwein1995adaptive}. 
In other words, this threshold is considered to be the difference between two stimuli intensities that can be recognized in some fixed percentage of the presentation (typically $50~\%$) by participants. Among different methods that can be used to determine the JND threshold, the \textit{adaptive staircase} method offers many advantages including higher efficiency, greater flexibility, and fewer assumptions (c.f. \cite{levitt1971transformed} for a detailed discussion). In order to identify exact threshold values with this method, it is necessary to be able to change the intensity of the stimuli in small steps.

The Speech-to-Noise Ratio (SNR) is the ratio of the power of the speech signal to the power of the background noise \cite{mcshefferty2015just}. 
The SNR directly affects perceived quality in telecommunication, as a poor SNR leads to reduced intelligibility \cite{mcshefferty2015just}. 
In order to determine the the JND, we thus used the SNR to change the intensity of stimuli (and as the result their quality) in small steps. 
Four speech samples (2 male, 2 female, German language) from the standard ITU-T Rec. P.501 \cite{ITU-P501} dataset were degraded with white additive noise to create a test dataset with different SNR levels ranging from 35dB to 50dB, with 1dB steps.

To determine the JND as a function of the SNR, an adaptive staircase psychoacoustics method (3AFC, 2~down--1~up) as proposed by Levit \cite{levitt1992adaptive} was implemented. 
During the test, pairs of stimuli (from a same source) are presented to the participant who should select which one has a better quality, or state that the "difference is not detectable", following a 3 Alternatives Forced Choice (3AFC) paradigm (c.f. Fig \ref{fig_pair_cmp}). 
The procedure starts from a JND of 15~dB SNR, in which one of the pairs is a stimulus with constant 50~dB~SNR (i.e. “reference” stimulus) and the other stimulus with 35~dB SNR (i.e. a “dynamic” stimulus).
When the participant correctly selects the stimulus with a better quality two times in a row for the current SNR, the SNR of the dynamic stimulus will increase by one step for the next question. However, by any wrong answer, the SNR of the next dynamic stimulus will be decreased again (i.e. 2~down--1~up). A reversal happens when the direction of movement of the dynamic stimulus changes (e.g. previously increased and now decrease due to a wrong answer or vice versa).
The test continues until either 7 reversals are observed (as suggested by Levit \cite{levitt1992adaptive}) or until the number of trials reached 45. In each reversal, the SNR in which the turn is happening is stored to calculate the JND level. 
The final resulting JND threshold is calculated by averaging between the SNR levels in last 6 reversals. This threshold represents the estimate of the SNR level for which in more than $70.7\%$ of times its difference with a reference stimulus is recognizable. In other words, for participant $p$ the JND of SNR is
\begin{align}
    JND_p = 50 - T_p, && T_p = \frac{1}{N-1} \sum_{i=2}^N t_{i,p}
\end{align}
where $t_{i,p}$ is the SNR in which the $i^{th}$ turn happened for the participant $p$ and $N$ is the number of reversals for that participant in the test (maximum is 7).
The estimated $JND_p$ depends to the hearing ability of the participant $p$, listening environment (namely background noise in the environment), and the listening system used by the participant $p$.

\subsection{Proposed Modified Test Method for Crowdsourcing} \label{mjndq}
We estimate the average JND in quality which normal-hearing participants can reach in a standard laboratory environment, according to ITU-T Rec. P.800, using an appropriate listening system.
When a crowdsourcing participant reaches the same JND threshold in their work environment using own listening hardware, we can reasonably assume that the participant’s environment (and test setup) is appropriate for performing a speech quality assessment test at that point of time. 
However, using an adaptive staircase test in the crowdsourcing session is not appropriate as it typically takes about 10 minutes and in crowdsourcing tasks are typically a lot shorter.
Therefore, we propose to ask participants to just rate four pairs representing the target JND in quality, which was on average achieved in the laboratory experiment, using the 3AFC paradigm. If the participant correctly answers \(\frac{3}{4}\) of the questions, he/she successfully passes the Environment Suitability Test.

\subsection{Laboratory Experiment}
We invited group of normal-hearing participants for a laboratory experiment. 
In each session, a participant performed the JNDQ test with three different listening devices (loudspeaker, own headphone, laboratory headphone) in two environments (silent room, and a simulation of 50dB(A) noise) in a randomized order. 
At the end, the hearing ability of each participant was measured with a professional audiometer \sebastian{What type of noise was used?}.
The environment was set up according to ITU-T Rec. P.800 in the silent case, and background noise was simulated according to ETSI ES 202 396-1 \cite{etsi202} in the corresponding noisy case. We used four speakers at the distance of 2m from participant to play the “Outside Traffic Street Noise” recorded at pavement level L 69.1dB(A) and R: 69.6dB(A) (c.f.  \cite{etsi202} for more details).
The Sound Pressure Level at the participant’s position was measured with an Artificial Head Measurement System\footnote{HMS II.3 from HEAD acoustics}. 
We decided to simulate 50dB(A) SPL environmental noise on participant's position, as 1) it is recommended as the highest level by ITU-T Rec. P.808, 2) \cite{jimenez2019annonym} showed that at such a level significant differences in quality ratings were observed compared to a quiet laboratory study.
The close-to-ear measurements of SPL by HMS II.3 during simulation of environmental noise showed 50.2~dB(A) SPL in channel 1, and 49.9~dB(A) in channel 2. When the laboratory headset\footnote{AKG 271} was used the close-to-ear measurements of SPL resulted to 42.6~dB(A) in channel 1 and 38.63~dB(A) in channel 2.
The results of the experiment are reported in the next section.

\subsection{Crowdsourcing Test}
Based on the result of the laboratory experiment, we examined three thresholds for the modified JNDQ test in crowdsourcing (c.f. section \ref{mjndq}).
The crowdsourcing tests were conducted according to ITU-T Rec. P.808. In each test, participants rated the speech quality of twelve degradation conditions represented each by 6 stimuli. 
The conditions and stimuli were taken from standard dataset 401 created according to ITU-T Rec. P.800. 
It was part of the dataset pool of the ITU-T Rec. P.863~\cite{ITU-P863} competition and was kindly shared with us for evaluating ITU-T Rec. P.808.

We selected the twelve degradation conditions that were constantly included in all datasets used in the ITU-T Rec. P.863~\cite{ITU-P863} competition. They cover the entire range of MOS, and include background noise and sub-optimum presentation level for which previous studies showed that the quality of those conditions would be rated differently in the presence of environmental noise compared to a quiet laboratory environment \cite{naderi2018speech}. 
A list of degradation conditions is provided in Table \ref{tab:deg}. 

In the crowdsourcing test, workers were first exposed to the hearing test and one variation of the modified JNDQ test, then they took part in the training and performed the rating task.
In the training section, they listened to 6 samples from the original dataset covering the entire range of MOS values. 
The aim of the training section is to help workers to familiarize with the system, and to also anchor their perception of quality for the stimuli under test to the scale range offered by the rating scale. 
Overall, in each crowdsourcing test, we aimed to collect 14 votes per file from crowdworkers.
\begin{table}[b]
\caption{Degradation conditions which were rated in the crowdsourcing tests, taken from dataset 401 of ITU-T Rec. P.863 competition pool.}
\label{tab:deg} 
\begin{center}
\begin{tabular}{cl}
\toprule
\textbf{Num} & \textbf{Degradation condition} \\
\midrule
C01 & SWB - Reference (clean), 0dB attenuation(14kHz)   \\
C02 & SWB - MNRU 10 dB\\
C03 & SWB - MNRU 20 dB\\
C04 & SWB - Background noise, 12dB Hoth SNR \\ 
C05 & SWB - Background noise, 20dB Babble SNR \\ 
C06 & SWB - Level acc. to P.56, -10 dB  \\
C07 & SWB - Level acc. to P.56, -20 dB \\
C08 & NB - MIRS-TX + MIRS-RX \\
C09 & NB - Band pass 500-2500 Hz\\
C10 & WB - Band pass 100-5000 Hz \\
C11 & SWB - Temporal clipping - 2\% packet loss \\
C12 & SWB - Temporal clipping - 20\% packet loss\\
\bottomrule
\end{tabular}
\end{center}
\end{table}

\section{Results}
\subsection{Laboratory experiment}
22 participants took part in the laboratory experiment. 
During the data cleansing procedure, 2 participants were removed as their answers failed in a reliability test. 
In addition, the responses of one more participant were removed due to the fact that the requirement of normal hearing ability was not observed in the audiometry test. 
As a result, the JND levels archived in different types of environment and using different devices for 19 participants were compared using factorial repeated-measure ANOVA.

The assumptions of sphericity were tested by Mauchly's test. 
It indicates that the assumptions had been violated for the interaction effect between the environment type and the device type, $\chi^2(2) = 7.549$, $p = .023$. 
Therefore, the degree of freedom for the interaction effect was corrected using a Greenhouse-Geissser estimate of sphericity ($\varepsilon = .736$).

\begin{figure}[htbp] 
  \centering
  \begin{subfigure}[b]{0.5\textwidth}
    \centering
    \includegraphics[width=\textwidth]{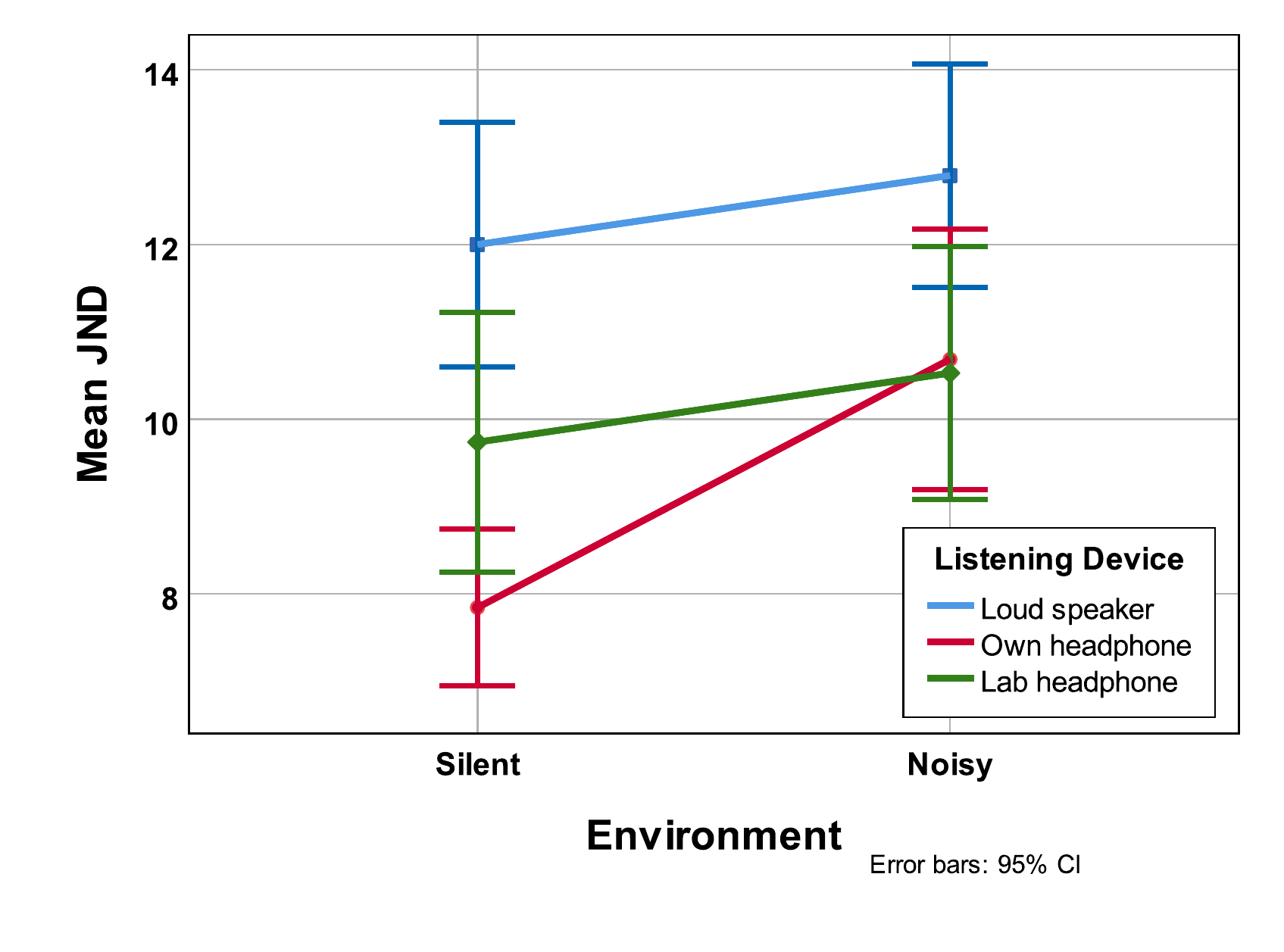} 
     \caption{Interaction graph between the environment types and the listening device types.} 
    \label{fig:annova_interaction} 
  \end{subfigure} 

  \begin{subfigure}[b]{0.5\textwidth}
    \centering
    \includegraphics[width=\textwidth]{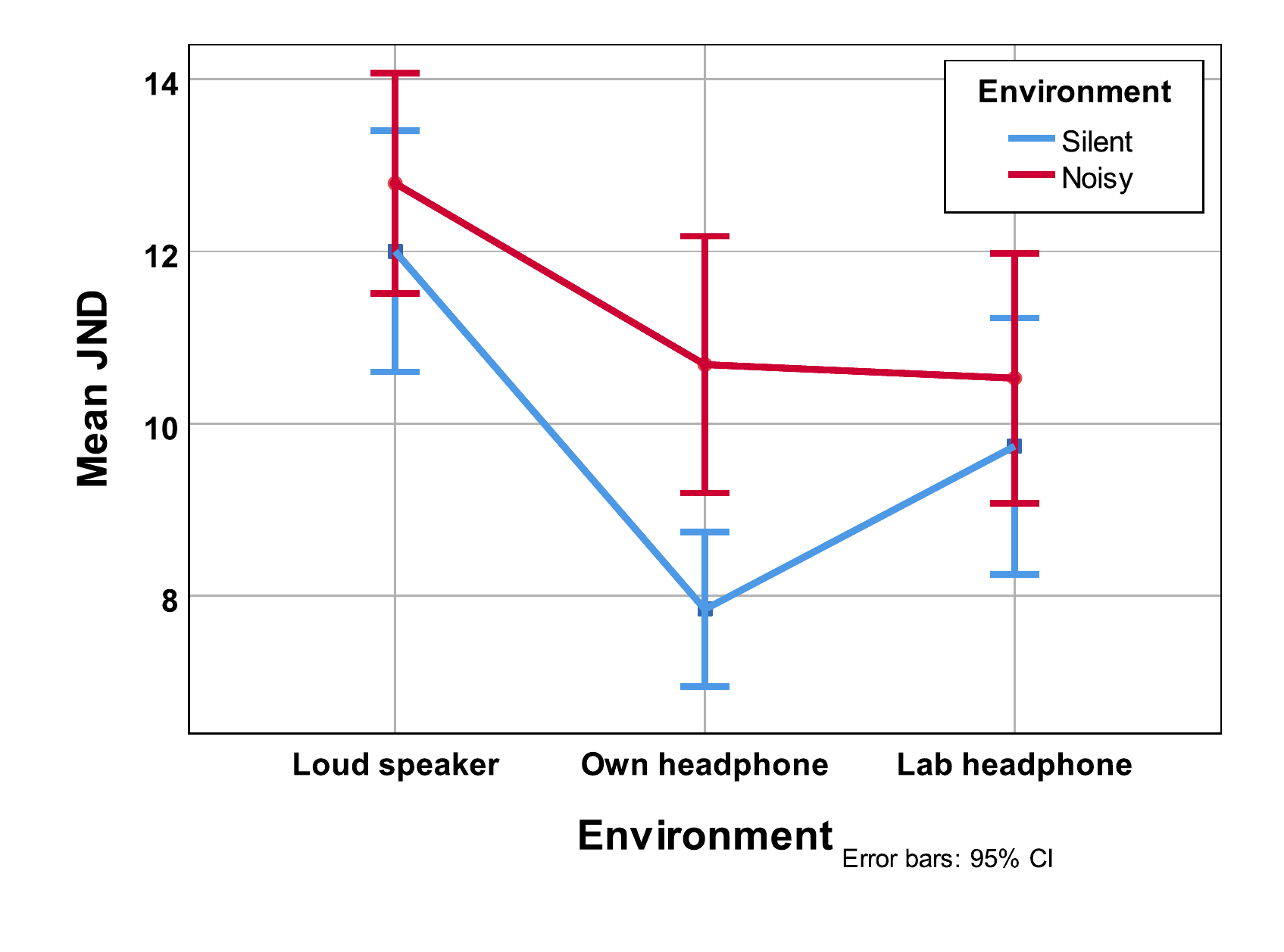}
    \caption{Mean JND in SNR achieved using different listening devices.}
	\label{fig_jnd_device} 
  \end{subfigure} 
  \caption{Result of factorial repeated-measure ANOVA for laboratory experiment.}
\end{figure}

There was a significant main effect of the type of environment on the JND level $F(1,18) = 13.74$, $p = .002$, $\eta^2 = .433$. 
Post-hoc test using the Bonferroni correction revealed that JND levels achieved in the silent environment ($M = 9.86$, $SEM= .50$) were significantly lower than levels achieved in the noisy environment ($M = 11.33$, $SEM= .56$), $p = .002$.
There was also a significant main effect of the type of listening device on the level of JND achieved $F(2, 36) = 21.8$, $p < .001$, $\eta^2 = .548$. 
Post-hoc test using the Bonferroni correction showed that there were statistically significant differences between JND level achieved with loudspeaker ($M=12.4$, $SEM=.55$) and own headphone ($M=9.26$, $SEM=.51$), $p < .001$, and also between loudspeaker and laboratory headphone ($M=10.13$, $SEM=.64$), $p =.001$. 
However, there was no significant difference between the level of JND achieved by own or laboratory headphones, $p=.369$. 
Figure \ref{fig_jnd_device} illustrates the mean JND level of SNR achieved using different listening devices.


There was also a significant interaction effect between the type of environment and the type of device used, $F(2,36) = 4.48$, $p = .031$, $\eta^2=.199$. 
This indicates that the type of device used had a differing effect on the level of JND depending on how noisy the environment was. 
Contrast \sebastian{I don't know what you mean by ``contrast''}\babak{It is statistical term used in ANNOVA, check wiki. I put a link but it mess up with latex ;)  } revealed a significant interaction when comparing own and laboratory headphones in different environments, $F (1,18) = 11$, $p= .004$, $r = .62$.
Figure \ref{fig:annova_interaction} shows that own and laboratory headphones might perform differently in the silent environment but similarly in the noisy environment.
 

\begin{figure}[htbp]
\centerline{\includegraphics[width=0.8\columnwidth]{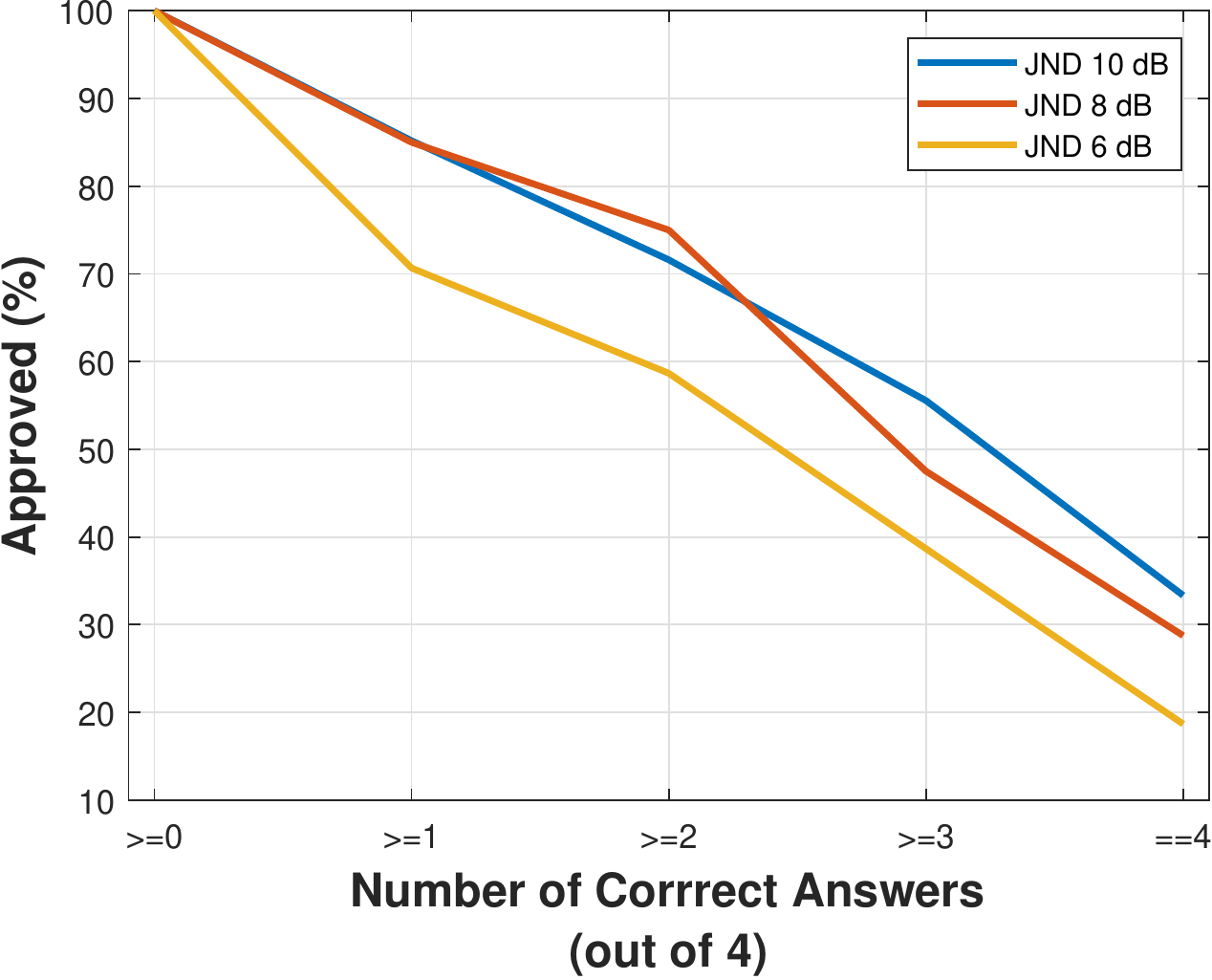}}
\caption{Percentage of answers approved by modified JNDQ test with different SNR level and acceptance criteria.}
\label{fig:cs:approved}
\end{figure}

\begin{table}[t]
\caption{Comparison between MOS values from laboratory experiment and MOS values calculated using ratings of submissions passed or failed in different modified JNDQ tests.}
\label{tab:performance} 
\begin{center}
\begin{tabular}{ccp{4mm}p{4mm}p{4mm}p{4mm}p{4mm}p{4mm}}
\toprule
\textbf{JND in} & \textbf{Acceptance} & \multicolumn{2}{c}{\textbf{PCC}}    & \multicolumn{2}{c}{\textbf{SRCC}} & \multicolumn{2}{c}{\textbf{RMSE}} \\
\textbf{SNR} & \textbf{criterion} &    \textit{Passed} & \textit{Failed} &  \textit{Passed} & \textit{Failed} &  \textit{Passed} & \textit{Failed} \\
\midrule
10 &	1/4	&\textbf{.968}\textsuperscript{a} &	\textbf{.751} &	\textbf{.958}\textsuperscript{a}	&\textbf{.732}&	\textbf{.326}\textsuperscript{a}	&\textbf{.758}\\
10 &	2/4	&.963&	.912&	\textbf{.958}\textsuperscript{b}	&\textbf{.860}&	.364	&.449\\
10 &	3/4	&.958&	.947&	.923	&.895&	.352	&.407\\
\midrule
8 & 	1/4	&.965&	.938&	.909	&.902&	.298	&.396\\
8 & 	2/4	&.965&	.937&	.909	&.846&	.298	&.406\\
8 & 	3/4	&.953&	.955&	.916	&.846&	.353	&.349\\
\midrule
6 & 	1/4	&.962&	.931&	.902	&.894&	.387	&.413\\
6 & 	2/4	&\textbf{.972}\textsuperscript{b}&	\textbf{.905}&	.923	&.890&	.350	&.491\\
6 & 	3/4	&\textbf{.980}\textsuperscript{b}&	\textbf{.920}&	\textbf{.965}\textsuperscript{a}	&\textbf{.846}&	\textbf{.301}\textsuperscript{b}	&\textbf{.480}\\

\bottomrule
\multicolumn{2}{l}{$^{a}$ Significant at $\alpha=.05$} & \multicolumn{6}{l}{$^{b}$ Significant at  $\alpha = .1$}
\end{tabular}
\end{center}
\end{table}

\begin{figure*}[h]
\centerline{\includegraphics[width=\textwidth]{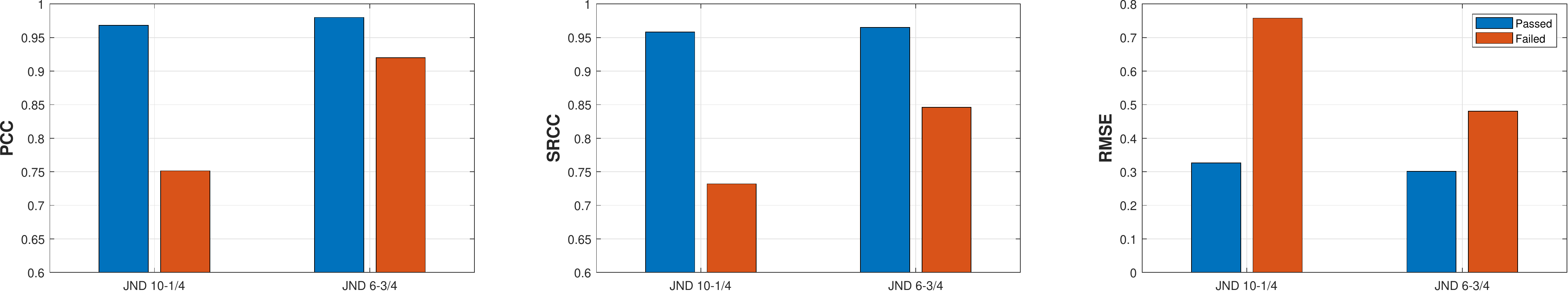}}
\caption{Comparison of MOS values in laboratory and MOS values calculated with ratings from passed and failed answers (only most lenient and strict setups).}
\label{fig:cs:bar_best}
\end{figure*}

\subsection{Crowdsourcing evaluation}
Based on the laboratory experiment, we selected three JND in SNR levels for the crowdsourcing evaluation, namely 10, 8 and 6 dB. 
\sebastian{I don't know why you chose these levels; in the noisy lab environment, the JND was always above 10 dB.}
\babak{I am looking for a test to find silent environments, and want to see how ratings of people passed or failed differs. Looking at Fig2 they seam to be reasonable thresholds }
The aim of the crowdsourcing evaluation was to observe if there is a difference between opinion ratings from participants who pass or fail in the modified JNDQ test with above-mentioned thresholds.
We selected 12 degradation conditions, and 6 stimuli per condition to be rated.
As number of files are higher than recommended number of stimuli per a crowdsourcing session \cite{ITU-P808}, we divided each test into six sessions (assignments).
In each assignment, participants rated 12 stimuli, one trapping question \cite{naderi2018motivation}, and one gold standard question. 
We included trapping questions and gold standard questions as recommended in \cite{ITU-P808} to check the reliability of the collected data. 
We used an open-source software package\urlfootnote{https://github.com/microsoft/P.808} to conduct the experiment in Amazon Mechanical Turk\urlfootnote{https://www.mturk.com}.
We conducted three crowdsourcing tests, each include one variation of modified JNDQ test (i.e. different JND in SNR level).
In each test, we aimed to collect 14 votes for each stimulus.
Overall 252 assignments (84 for each test) were posted for crowd workers.
Among 288\footnote{There is a difference between create assignments and submitted answers because we actively published a new assignment when a submitted answer failed in quality control step.} submitted answers, 52 responses were removed in the data cleansing step (because of wrong usage of headphone, wrong answer to the trapping question or invalid answer to the gold standard question).
Each crowdsourcing test (i.e. JND group) has nearly similar number of answers ($M = 79$,  $STD= 3.2$), which passed the first data cleansing step, overall submitted by $93$ crowd workers.

In each test, different percentage of answers pass the modified JNDQ test depending to the SNR level used and the acceptance criteria (i.e. minimum number of correct answers out of the four questions).
Figure \ref{fig:cs:approved} illustrates the  percentage of answers passed the modified JNDQ test based on the JND in SNR levels and different acceptance criteria.
Since setting the criteria in such a way that all of the 4 pair-comparison questions should be correctly answered leads to failed of more than $60\%$ of assignments, we do not consider this criterion in the next steps. 
We divided the submitted answers into two groups; "Passed" answers which passed the corresponding modified JNDQ test and "Failed" answers which failed the test.
We also considered different criterion (i.e. number of questions correctly answered out of four) to decide whether the test is passed or failed.
For each group we calculated MOS per 12 conditions and the Pearson Correlation Coefficient (PCC), Spearman's Rank correlation coefficient (SRCC) and the Root Mean Square Error (RMSE) between the calculated MOS and the MOS values from the standard laboratory experiment performed according to the ITU-T Rec. P.800.
Results are presented in Table\ref{tab:performance}. 
The highest correlation to the laboratory achieved in the most strict test setup i.e. JND in SNR level of $6$ and three or more questions out of four should be answered correctly. 
Considering the RMSE, the least value achieved with JND of $8$ and 1 or more correct answers.
We also compared the difference between correlation coefficients and RMSE values from the passed and the failed groups using Fisher-z transformation and F-distribution respectively, as suggested in \cite{ITU-P1401}. 
Results show significant improvement in both most strict and most lenient setups. 
Figure \ref{fig:cs:bar_best} illustrates those cases.

\begin{figure*}[htbp]
\centerline{\includegraphics[width=\linewidth]{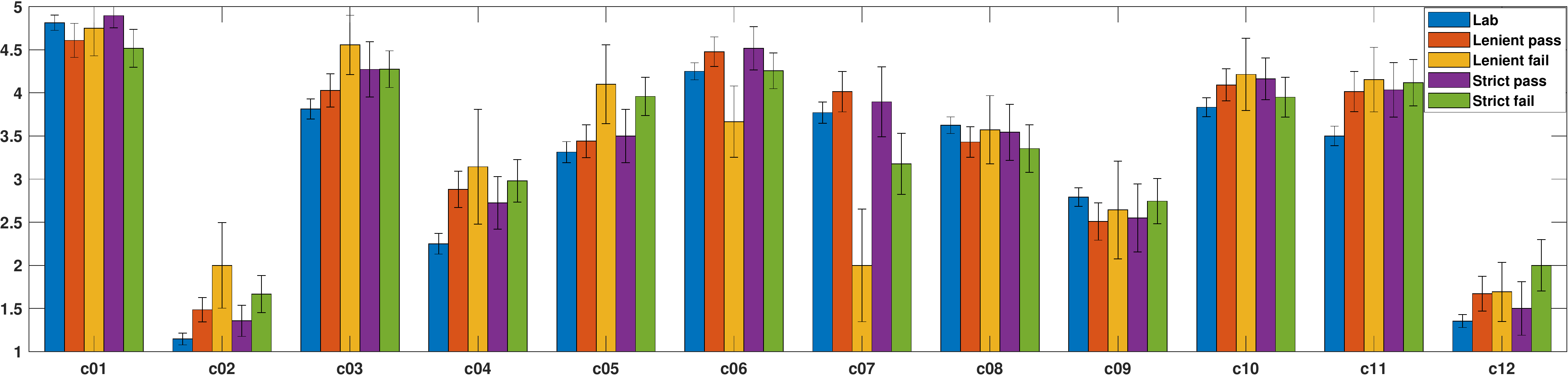}}
\caption{MOS values per degradation conditions, provided by each group.}
\label{fig:cs:cond}
\end{figure*}

We also evaluated how MOS for degradation conditions provided by the passed and failed groups differ for the strict and lenient setups (cf. Figure \ref{fig:cs:cond}).
Conditions C01, C08 and C09 (i.e. reference and narrow-bands) are rated similarly by passed and failed groups and very close to the laboratory MOS.
Crowd workers provided higher rating for C10, C11 and C12 than laboratory MOS (constant 
between all groups).
For conditions C02-5 (i.e. noise) always both failed groups gave higher ratings and specifically people failed in the lenient setup.
For condition C06 and C07 (i.e. sub-optimum presentation level), both failed groups gave a lower quality ratings. Again specifically ratings of group which failed in the lenient setup was strongly lower than laboratory MOS values. 
These results are inline with finding in the literature \cite{naderi2018speech} where in the laboratory environment background noise in different levels were simulated and participants performed speech quality assessment test.

\section{Discussion and conclusion}
In this paper we assessed the application of JND in quality as an environment suitability test for crowdsourcing. 
Laboratory and crowdsourcing experiments were conducted for this purpose.
Our results show that the achievable JND level in a laboratory depends on the background noise and on the listening device, with an interaction effect between them.
As a consequence, a properly designed JND test seems to be appropriate for distinguishing noisy environment from silent conditions, once the listening device is known.
In addition, the JND seems to be particularly high when using loudspeakers.
As listening through loudspeakers are not recommended for speech quality assessment\cite{ITU-P808}, a well-designed JND test may also be used for filtering out participants who do not use a headphone in crowdsourcing tests\footnote{To the best of authors' knowledge there is no comprehensive way to detect usage of headphones utilizing software libraries.}.

We evaluated the modified JNDQ test in a crowdsourcing experiment conducted according to the ITU-T Rec. P.808. 
Participants were first exposed to a hearing test, and then training and rating sections. 
We created three environment tests (i.e. a modified JNDQ test with 4 questions all representing a same JND in SNR level). 
An environment test was always positioned before the rating section.
In each environment test we took a different JND in SNR level namely 10, 8 and 6 dB. 
Despite passing or failing in the environment test, participants got access to the ratings task.
In more than $60\%$ of times, participants failed in at least 1/4 questions, with higher percentage of failures in lower JND levels.
The ratings provided after passing or failing in each environment test were compared to the ratings provided in a standard laboratory test.
Significant difference between performance of passed and failed groups, in term of correlation of their votes to the laboratory MOS values, were observed in the most strict (i.e. JND 6 dB and +3/4 correct answer) and most lenient (i.e. JND 10 dB and +1/4 correct answer) setups.
Our studies show that the lenient case is more cost efficient as there only $15\%$ of answers were failed, whereas the failed ratio in the strict case is $61\%$.
In addition, detailed look on the MOS ratings for each degradation condition, revealed that applying the modified JNDQ test leads to reliable assessment of degradation conditions that include noise and sub-optimum presentation level.
It should be considered that since the modified JNDQ test evaluates the suitability of participant's environment on that specific time, it is not a continues monitoring solution.
On one hand, as the worker's surrounding acoustic scene can be changed frequently, it is recommended to expose them to the test often.
On the other hand, injecting the modified JNDQ test in every CS session strongly increases the overall duration of the session. 
Finding an efficient rate in which the modified JNDQ test should be integrated in a crowdsourcing session is subject of further research.
Overall, we recommend to use the lenient setup to filter out participants with inappropriate setup.
Depending on the goal of experiment, one may consider on how often include the test and to use a more strict configuration.

\bibliographystyle{IEEEtran}
\bibliography{library}

\begin{thebibliography}{10}
\providecommand{\url}[1]{#1}
\csname url@samestyle\endcsname
\providecommand{\newblock}{\relax}
\providecommand{\bibinfo}[2]{#2}
\providecommand{\BIBentrySTDinterwordspacing}{\spaceskip=0pt\relax}
\providecommand{\BIBentryALTinterwordstretchfactor}{4}
\providecommand{\BIBentryALTinterwordspacing}{\spaceskip=\fontdimen2\font plus
\BIBentryALTinterwordstretchfactor\fontdimen3\font minus
  \fontdimen4\font\relax}
\providecommand{\BIBforeignlanguage}[2]{{%
\expandafter\ifx\csname l@#1\endcsname\relax
\typeout{** WARNING: IEEEtran.bst: No hyphenation pattern has been}%
\typeout{** loaded for the language `#1'. Using the pattern for}%
\typeout{** the default language instead.}%
\else
\language=\csname l@#1\endcsname
\fi
#2}}
\providecommand{\BIBdecl}{\relax}
\BIBdecl

\bibitem{le2012qualinet}
P.~Le~Callet, S.~M{\"o}ller, A.~Perkis \emph{et~al.}, ``Qualinet white paper on
  definitions of quality of experience,'' \emph{European network on quality of
  experience in multimedia systems and services (COST Action IC 1003)}, vol.~3,
  no. 2012, 2012.

\bibitem{Jekosch05}
U.~Jekosch, \emph{Voice and Speech Quality Perception. Assessment and
  Evaluation}.\hskip 1em plus 0.5em minus 0.4em\relax Berlin: Springer, 2005.

\bibitem{ITU-P800}
{ITU-T Rec. P.800}, \emph{{Methods for subjective determination of transmission
  quality}}.\hskip 1em plus 0.5em minus 0.4em\relax Geneva: International
  Telecommunication Union, 1996.

\bibitem{ITU-P808}
{ITU-T Rec. P.808}, \emph{{Subjective evaluation of speech quality with a
  crowdsourcing approach}}.\hskip 1em plus 0.5em minus 0.4em\relax Geneva:
  International Telecommunication Union, 2018.

\bibitem{naderi2018motivation}
B.~Naderi, \emph{Motivation of workers on microtask crowdsourcing
  platforms}.\hskip 1em plus 0.5em minus 0.4em\relax Springer, 2018.

\bibitem{jimenez2019background}
R.~Z. Jim{\'e}nez, B.~Naderi, and S.~M{\"o}ller, ``Background environment
  characteristics of crowd-workers from german speaking countries experimental
  survey on user environment characteristics,'' in \emph{QoMEX 2019}.\hskip 1em
  plus 0.5em minus 0.4em\relax IEEE, 2019, pp. 1--3.

\bibitem{ribeiro2011crowdsourcing}
F.~Ribeiro, D.~Florencio, and V.~Nascimento, ``Crowdsourcing subjective image
  quality evaluation,'' in \emph{2011 18th IEEE International Conference on
  Image Processing}.\hskip 1em plus 0.5em minus 0.4em\relax IEEE, 2011, pp.
  3097--3100.

\bibitem{naderi2018speech}
B.~Naderi, S.~M{\"o}ller, and G.~Mittag, ``Speech quality assessment in
  crowdsourcing: Influence of environmental noise,'' \emph{DAGA}, pp. 299--302,
  2018.

\bibitem{treutwein1995adaptive}
B.~Treutwein, ``Adaptive psychophysical procedures,'' \emph{Vision research},
  vol.~35, no.~17, pp. 2503--2522, 1995.

\bibitem{levitt1971transformed}
H.~Levitt, ``Transformed up-down methods in psychoacoustics,'' \emph{The
  Journal of the Acoustical society of America}, vol.~49, no.~2B, pp. 467--477,
  1971.

\bibitem{mcshefferty2015just}
D.~McShefferty, W.~M. Whitmer, and M.~A. Akeroyd, ``The just-noticeable
  difference in speech-to-noise ratio,'' \emph{Trends in hearing}, vol.~19,
  2015.

\bibitem{ITU-P501}
{ITU-T Recommendation P.501}, \emph{{Test signals for use in
  telephonometry}}.\hskip 1em plus 0.5em minus 0.4em\relax Geneva:
  International Telecommunication Union, 2017.

\bibitem{levitt1992adaptive}
H.~Levitt, ``Adaptive procedures for hearing aid prescription and other
  audiologic applications,'' \emph{Journal of the American Academy of
  Audiology}, vol.~3, no.~2, pp. 119--131, 1992.

\bibitem{etsi202}
{ETSI 202 396-1}, \emph{{Speech quality performance in the presence of
  background noise; Part 1: Background noise simulation technique and
  background noise database}}.\hskip 1em plus 0.5em minus 0.4em\relax Speech
  and multimedia Transmission Quality (STQ), 2017.

\bibitem{jimenez2019annonym}
R.~Z. Jim{\'e}nez, B.~Naderi, and S.~M{\"o}ller, ``Effect of environmental
  noise in speech quality assessment studies using crowdsourcing,'' in
  \emph{QoMEX 2020}.\hskip 1em plus 0.5em minus 0.4em\relax IEEE, 2020.

\bibitem{ITU-P863}
{ITU-T Rec. P.863}, \emph{{Perceptual objective listening quality
  prediction}}.\hskip 1em plus 0.5em minus 0.4em\relax Geneva: International
  Telecommunication Union, 2018.

\bibitem{ITU-P1401}
{ITU-T Rec. P.1401}, \emph{{Methods, metrics and procedures for statistical
  evaluation, qualification and comparison of objective quality prediction
  models}}.\hskip 1em plus 0.5em minus 0.4em\relax Geneva: International
  Telecommunication Union, 2012.

\end{thebibliography}

\end{document}